\documentstyle[epsfig]{elsart}
\begin{document}
\begin{frontmatter}
\title{Multifractality of river runoff and precipitation: Comparison of 
fluctuation analysis and wavelet methods}
\author[gi]{Jan W. Kantelhardt}, \author[gi]{Diego Rybski},
\author[ma,gi]{Stephan A. Zschiegner}, \author[mu]{Peter Braun}, 
\author[gi,pik]{Eva Koscielny-Bunde}, \author[il]{Valerie Livina}, 
\author[il]{Shlomo Havlin}, and \author[gi]{Armin Bunde}
\address[gi]{Institut f\"ur Theoretische Physik III,
Justus-Liebig-Universit\"at, Giessen, Germany}
\address[ma]{Klinik f\"ur Innere Medizin, Klinikum der
Philipps-Universit\"at, Marburg, Germany}
\address[mu]{Bayerisches Landesamt f\"ur Wasserwirtschaft, M\"unchen, 
Germany}
\address[pik]{Institute for Climate Impact Research, Potsdam, Germany}
\address[il]{Dept. of Physics and Minerva Center, Bar-Ilan University, 
Ramat-Gan, Israel}
\date{April 5, 2003}

\begin{abstract}
We study the multifractal temporal scaling properties of river discharge 
and precipitation records.  We compare the results for the 
multifractal detrended fluctuation analysis method with the results for 
the wavelet transform modulus maxima technique and obtain agreement within
the error margins.  In contrast to previous studies, we find non-universal 
behaviour:  On long time scales, above a crossover time scale of several 
months, the runoff records are described by fluctuation exponents varying 
from river to river in a wide range.  Similar variations are observed for
the precipitation records which exhibit weaker, but still significant 
multifractality.  For all runoff records the type of multifractality 
is consistent with a modified version of the binomial multifractal model, 
while several precipitation records seem to require different models.
\end{abstract}\end{frontmatter}

The analysis of river flows has a long history.  Already more than half 
a century ago the engineer H. E. Hurst found that runoff records from various 
rivers exhibit 'long-range statistical dependencies' \cite{Hurst51}.  
Later, such long-term correlated fluctuation behaviour has also been 
reported for many other geophysical records including precipitation data
\cite{Hurst65,Mandelbrot69}, see also \cite{Feder88}.  These original 
approaches exclusively focused on the absolute values or the variances of 
the full distribution of the fluctuations, which can be regarded as the 
first moment $F_1(s)$ \cite{Hurst51,Hurst65,Mandelbrot69} and the second 
moment $F_2(s)$ \cite{Matsoukas00}, respectively.  In the last decade 
it has been realized that a multifractal description is required for a 
full characterization of the runoff records \cite{Tessier96,Pandey98}.  
Accordingly, one has to consider all moments $F_q(s)$ to fully characterize 
the records.  This multifractal description of the records can be regarded 
as a 'fingerprint' for each station or river, which, among other things, can 
serve as an efficient non-trivial test bed for the state-of-the-art 
precipitation-runoff models. 

Since a multifractal analysis is not an easy task, especially if the data 
are affected by trends or other non-stationarities, e.g. due to a modification 
of the river bed by construction work or due to changing climate, it is 
useful to compare the results for different methods.  We have studied the 
multifractality by using the multifractal detrended fluctuation analysis 
(MF-DFA) method \cite{Kantelh02} (see also \cite{Kosciel98,Talkner01}) and 
the well established wavelet transform modulus maxima (WTMM) technique 
\cite{Muzy91,Arneodo02} and find that both methods yield equivalent results.  
Both approaches differ from the multifractal approach introduced into 
hydrology by Lovejoy and Schertzer \cite{Tessier96,Pandey98}.

\begin{figure} 
\begin{center} \epsfig{file=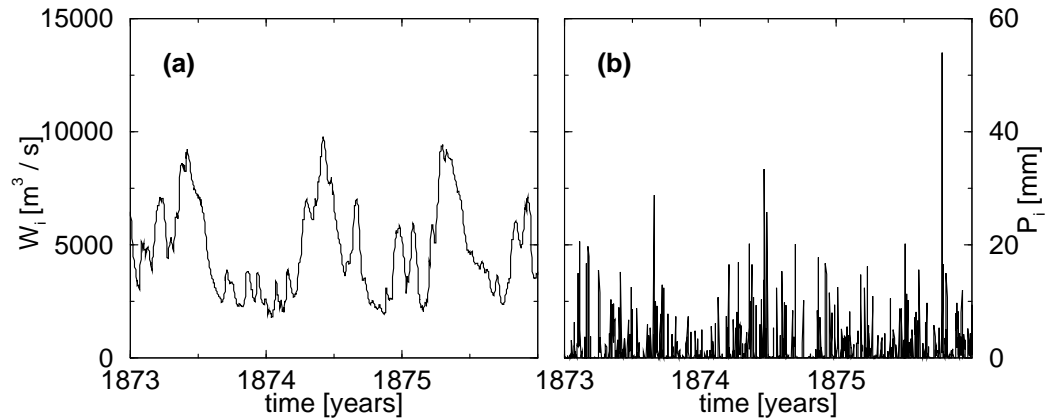,width=6cm,angle=-90} \end{center}
\caption{Three years of (a) the daily runoff record of the river Danube 
(Orsova, Romania) and (b) of the daily precipitation recorded in Vienna 
(Austria).} \label{fig:1}\end{figure}

We analyze long daily runoff records $\{W_i\}$ from six international 
hydrological stations and long daily precipitation records $\{P_i\}$ from 
six international meteorological stations.  The stations are representative 
for different rivers and different climate zones, as we showed in larger
separate studies \cite{WRR03,diego}.  As a representative example, Fig. 1 
shows three years of the runoff record of the river Danube (a) and of the 
precipitation recorded in Vienna (b).  It can be seen that the precipitation 
record appears more random than the runoff record.  To eliminate the periodic 
seasonal trend, we concentrate on the departures $\phi_i=W_i - \overline W_i$ 
(and $\phi_i=P_i - \overline P_i$) from the mean daily runoff $\overline W_i$.  
$\overline W_i$ is calculated for each calendar date $i$, e.g. 1st of April, 
by averaging over all years in the record.

\begin{figure} 
\begin{center} \epsfig{file=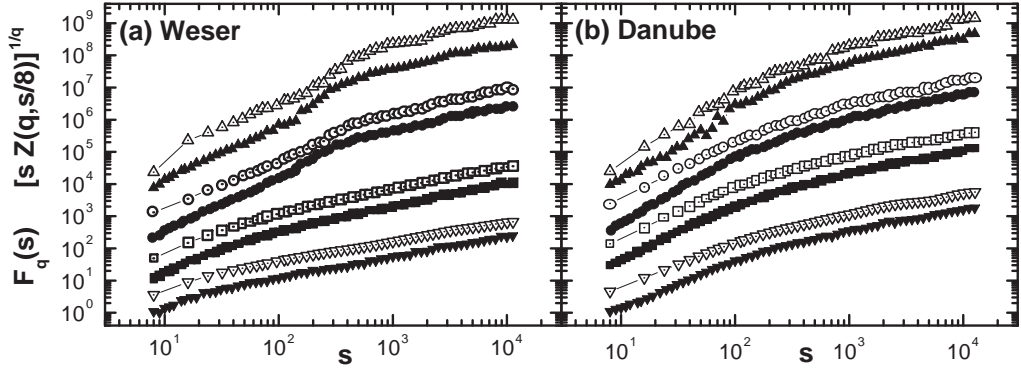,width=13.5cm} \end{center}
\caption{Comparison of the fluctuation functions $F_q(s)$ calculated with
the multifractal detrended fluctuation analysis (MF-DFA, filled symbols) 
with the rescaled wavelet transform modulus maxima (WTMM) partition sums 
$[s Z(q,s/8)]^{1/q}$ (open symbols) as function of time scale $s$ (in days)
for (a) the river Weser (Vlotho, Germany, 171y) and (b) the river Danube 
(Orsova, Romania, 151y).  The different symbols indicate different moments, 
$q=-6$ (triangles up), $q=-2$ (circles), $q=2$ (squares), $q=6$ (triangles 
down), and the curves are shifted vertically for clarity.  The slopes $h(q)$
for large $s$ of both, the MF-DFA curves and the rescaled WTMM curves are 
equivalent.} \label{fig:2}\end{figure}

In the MF-DFA procedure \cite{Kantelh02}, the moments $F_q(s)$ are calculated 
by (i) integrating the series, (ii) splitting the series into segments of 
length $s$, (iii) calculating the mean-square deviations $F^2(\nu,s)$ from 
polynomial fits in each segment, (iv) averaging $[F^2(\nu,s)]^{q/2}$ 
over all segments, and (v) taking the $q$th root.  In the paper, we have 
used third order polynomials in the fitting procedure of step (iii) (MF-DFA3), 
this way eliminating quadratic trends in the data.  We consider both, 
positive and negative moments $F_q(s)$ ($q$ ranges from $-10$ to $+10$) and 
determine them for time scales $s$ between $s=5$ and $s=N/5$, where $N$ is 
the length of the series.  Figure 2 shows the results (filled symbols) for 
two representative hydrological stations.  On large time scales, above a 
crossover occurring around 30-200 days, we observe a power-law scaling 
behaviour,
\begin{equation} F_q(s) \sim s^{h(q)}, \label{Fqs} \end{equation}
where the scaling exponent $h(q)$ (the slope in Fig. 2) explicitly depends 
on the value of $q$.  This behaviour represents the presence of 
multifractality.  

In order to test the MF-DFA approach we have applied the well-established 
WTMM technique, which is also detrending but based on wavelet analysis 
instead of polynomial fitting procedures.  For a full description of the 
method, we refer to \cite{Muzy91,Arneodo02}.  First, the  wavelet-transform 
$T(n,s') = {1 \over s'} \sum_{i=1}^N \phi_i \, g[(i-n)/s']$ of the 
departures $\phi_i$ is calculated.  For the wavelet $g(x)$ we choose 
the third derivative of a Gaussian here, $g(x)=d^3(e^{-x^2/2})/dx^3$, 
which is orthogonal to quadratic trends.  Now, for a given scale $s'$, 
one determines the positions $n_i$ of the local maxima of $\vert T(n,s') 
\vert$, so that $\vert T(n_i-1,s') \vert < \vert T(n_i,s') \vert \ge \vert 
T(n_i+1,s') \vert$.  Then, one obtains the WTMM partition sum $Z_q(s')$ by 
averaging $\vert T(n_i,s') \vert^q$ for all maxima $n_i$.  An additional 
supremum procedure has to be used in the WTMM method in order to keep the 
dependence of $Z(q,s')$ on $s'$ monotonous \cite{Arneodo02}.  The expected 
scaling behaviour is $Z(q,s') \sim (s')^{\tau(q)}$, where $\tau(q)$ are the 
Renyi exponents.  Since $\tau(q)$ is related to the exponents $h(q)$ by 
$h(q)= [\tau(q) + 1]/q$ \cite{Kantelh02} we have plotted 
\begin{equation} [s Z(q,s/8)]^{1/q} \sim s^{[\tau(q) + 1]/q} \sim s^{h(q)}. 
\label{Zqs}\end{equation}
We set $s' = s/8$ in the comparison with the MF-DFA results, since the 
wavelet we employ can be well approximated within a window of size $8s'$ 
(i.e. within 4 standard deviations on both sides), and this window size 
corresponds to the segment length $s$ in the MF-DFA.  Figure 2 shows that 
both methods yield equivalent results for the $q$ values we considered.

\begin{figure} 
\begin{center} \epsfig{file=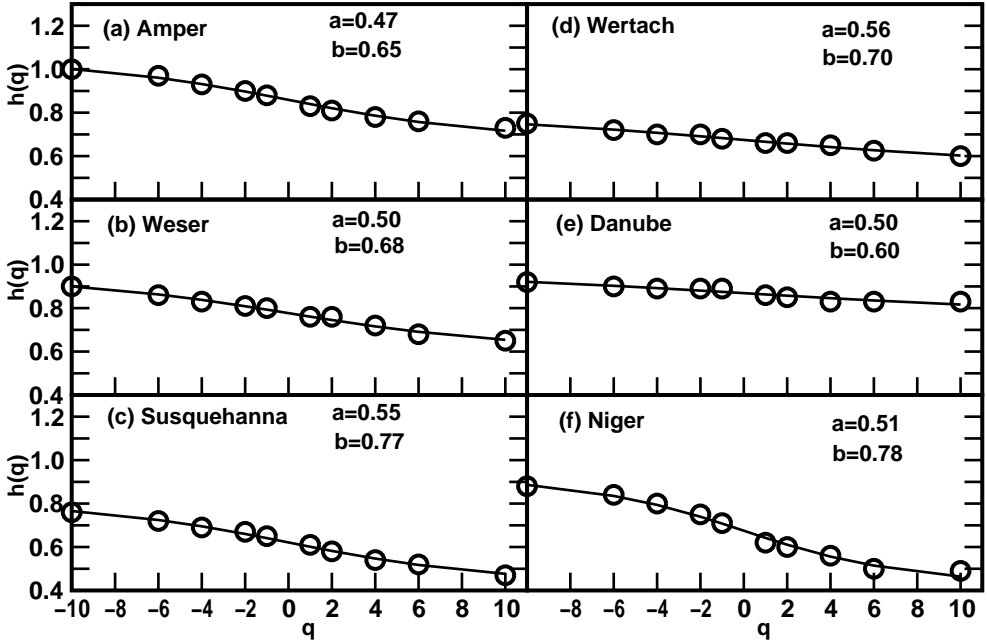,width=8.5cm,angle=-90} \end{center}
\caption{The generalized Hurst exponents $h(q)$ for six representative daily 
runoff records: (a) Amper in F\"urstenfeldbruck, Germany, (b) Weser in 
Vlotho, Germany, (c) Susquehanna in Harrisburg, USA, (d) Wertach in 
Biessenhofen, Germany, (e) Danube in Orsova, Romania, and (f) Niger in 
Koulikoro, Mali.  The $h(q)$ values have been determined by straight line 
fits of $F_q(s)$ on large time scales.  The error bars of the fits correspond 
to the size of the symbols.  The lines are obtained by fits of the 
two-parameter binomial model yielding Eq.~(\ref{bin}).  The resulting model 
parameters $a$ and $b$ are reported in the figures.  All fits are consistent 
with the data within the error bars (from \cite{WRR03}).} 
\label{fig:3}\end{figure}

Using the MF-DFA results, we have determined $h(q)$ from Eq.~(\ref{Fqs}) 
for all runoff records and all precipitation records and for several values 
of $q$.  Since a crossover occurs in $F_q(s)$ for time scales in the range 
of 30-200 days, we considered only sufficiently long time scales (above one 
year), where the results scale well.  Figure 3 shows $h(q)$ for the runoff 
data, while Fig. 4 shows $h(q)$ for the precipitation data.  Together with 
the results we show least-square fits according to the formula
\begin{equation} h(q)  = {1 \over q} - {\ln[a^q + b^q] \over q\ln 2},
\label{bin} \end{equation}
which corresponds to $\tau(q)=-\ln[a^q+b^q]/\ln 2$ and can be obtained 
from a generalized binomial multifractal model \cite{WRR03}, see also 
\cite{Feder88,Kantelh02}.  The values of the two parameters $a$ and $b$
are also reported in the figures.  The results for all rivers can be fitted 
surprisingly well with only these two parameters (see Fig. 3).  Instead of 
choosing $a$ and $b$, we could also choose the Hurst exponent $h(1)$ and the 
persistence exponent $h(2)$.  From knowledge of two moments, all the other
moments follow.

\begin{figure} 
\begin{center} \epsfig{file=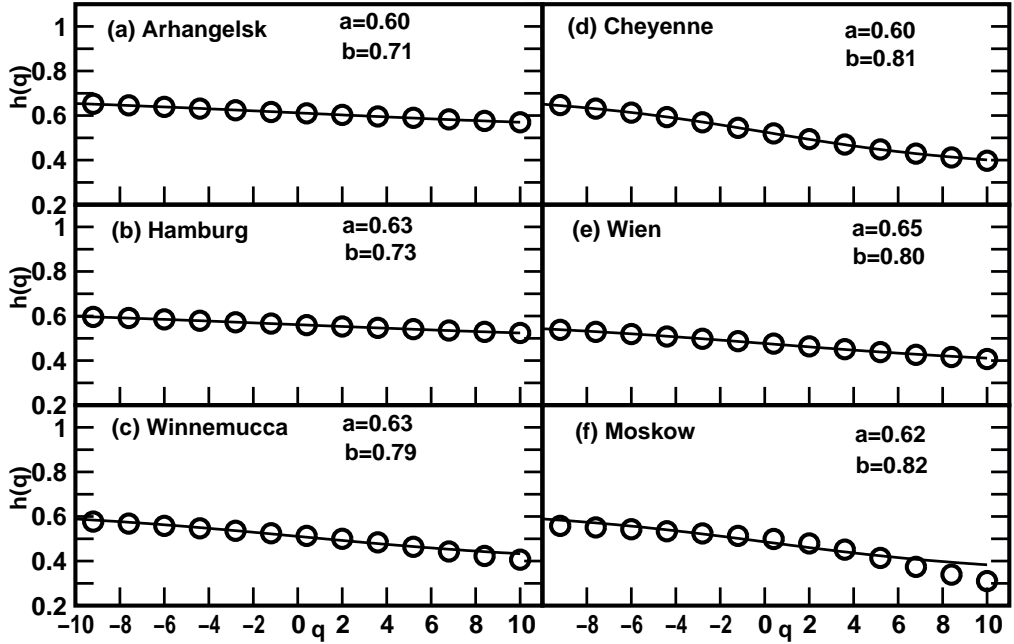,width=8.5cm,angle=-90} \end{center}
\caption{The generalized Hurst exponents $h(q)$ for six representative daily 
precipitation records:  (a) Arhangelsk (Russia), (b) Hamburg (Germany), 
(c) Winnemucca (USA), (d) Cheyenne (USA), (e) Vienna  (Austria), (f) Moskow  
(Russia), analogous with Fig. 3.  While the fits in (a,b,d,e) are consistent 
with the data within the error bars, significant deviations occur in (c) 
and -- even more drastically -- in (f).} \label{fig:4}\end{figure}

This surprising result does not hold for the precipitation records.
As can be seen in Figs. 4(c) and 4(f) there are stations where 
Eq.~(\ref{bin}) cannot describe the multifractal scaling behaviour 
reasonably well.  According to Rybski et al., Eq.~(\ref{bin}) is appropriate 
only for about 50 percent of the precipitation records \cite{diego}. 

In the generalized binomial multifractal model, the strength of 
multifractality is described by the difference of the asymptotical 
values of $h(q)$, $\Delta \alpha \equiv h(-\infty) - h(\infty) = (\ln b - 
\ln a) /\ln 2$.  We note that this parameter is identical to the width 
of the singularity spectrum $f(\alpha)$ at $f = 0$.  Studying 41 river 
runoff records \cite{WRR03}, we have obtained an average $\Delta \alpha = 
0.49 \pm 0.16$, which indicates rather strong multifractality on the long 
time scales considered.  For the precipitation records, on the other hand,
the multifractality is weaker.  The average is $\Delta \alpha = 0.29 
\pm 0.14$ for 83 records \cite{diego}.

Our results for $h(q)$ may be compared with the different ansatz $h(q) 
= 1 + H' - C_1 (q^{\alpha'-1} - 1) /(\alpha'-1)$ with the three 
parameters $H'$, $C_1$, and $\alpha'$ (LS ansatz), that has been used 
by Lovejoy, Schertzer, and coworkers \cite{Tessier96,Pandey98} successfully 
to describe the multifractal behaviour of rainfall and runoff records for 
$q>0$.  A quantitative comparison between both methods is 
inhibited, since here we considered only long time scales and used 
detrending methods.  We like to note that formula (\ref{bin}) for $h(q)$ 
is not only valid for positive $q$ values, but also for negative $q$ values.  
We find it remarkable, that for the runoff records only two parameters were 
needed to fit the data.  For the precipitation data, one needs either three
parameters like in the LS ansatz or different schemes.

In summary, we have analyzed long river discharge records and long
precipitation records using the multifractal detrended fluctuation 
analysis (MF-DFA) and the wavelet transform modulus maxima (WTMM) 
method.  We obtained agreement within the error margins and found that
the runoff records are characterized by stronger multifractality than
the precipitation records.  Surprisingly, the type of multifractality 
occurring in all runoff records is consistent with a modified version 
of the binomial multifractal model, which supports the idea of a 'universal' 
multifractal behaviour of river runoffs suggested by Lovejoy and Schertzer.
In contrast, according to \cite{diego}, several precipitation records 
seem to require a different description or a three-parameter fit like the 
LS ansatz.  The multifractal exponents can be regarded as 'fingerprints' 
for each station.  Furthermore, a multifractal generator based on the 
modified binomial multifractal model can be used to generate surrogate 
data with specific properties for each runoff record and for some of 
the precipitation records.  

{\it Acknowledgments:} We would like to thank the German Science 
Foundation (DFG), the German Federal Ministry of Education and Research
(BMBF), the Israel Science Foundation (ISF), and the Minerva Foundation 
for financial support.  We also would like to thank H. \"Osterle for 
providing some of the observational data.

\end{document}